\documentclass[aps,prl,twocolumn,nofootinbib,longbibliography,superscriptaddress]{revtex4-1}
\usepackage{hyperref}
\usepackage{amsmath}
\usepackage{amssymb}
\usepackage{amsthm}
\usepackage{graphicx}
\usepackage{color}
\newcommand{\be}{\begin{equation}}
\newcommand{\ee}{\end{equation}}

\newcommand{\bi}{\begin{itemize}}
\newcommand{\ei}{\end{itemize}}

\newcommand{\bfig}{\begin{figure}\begin{center}}
\newcommand{\efig}{\end{center}\end{figure}}

\def\ba#1\ea{\begin{align}#1\end{align}}
\def\bg#1\eg{\begin{gather}#1\end{gather}}

\def\({\left(}
\def\){\right)}
\def\[{\left[}
\def\]{\right]}
\def\<{\langle}
\def\>{\rangle}

\theoremstyle{definition}

\begin{document}

\title{Cobordism Conjecture in AdS}

{\small CALT-TH 2020-028,
IPMU 20-0073, YITP-20-84
\\
\\
\\
\\}

\author{Hirosi Ooguri}
\affiliation{Walter Burke Institute for Theoretical Physics, California Institute of Technology,  Pasadena, CA 91125, USA}
\affiliation{Kavli Institute for the Physics and Mathematics of the Universe (WPI), University of Tokyo,
   Kashiwa, 277-8583, Japan}
\author{Tadashi Takayanagi}
\affiliation{Yukawa Institute for Theoretical Physics,
Kyoto, 606-8502, Japan}
\affiliation{Kavli Institute for the Physics and Mathematics of the Universe (WPI), University of Tokyo,
   Kashiwa, 277-8583, Japan}
\affiliation{Inamori Research Institute for Sciences,
Kyoto, 600-8411, Japan}
\begin{abstract}
McNamara and Vafa conjectured that any pair of consistent quantum gravity theories can be connected by a domain wall. We test the  conjecture in the context of the AdS/CFT correspondence. 
There are topological constraints on existence of an interface between the corresponding conformal field theories. We discuss how to construct domain walls in AdS predicted by the conjecture 
when  the corresponding conformal interfaces are prohibited by topological obstructions. 
\end{abstract}

\maketitle
\vskip 1cm

In \cite{McNamara:2019rup}, McNamara and Vafa conjectured that
any proposed quantum theory of gravity with non-trivial cobordism classes in the space of configurations belongs to the Swampland. 
Their cobordism conjecture was motivated by the expectation that
string theory is a unique theory of quantum gravity rather than a collection of many different independent theories.
They argued that cobordism classes are  conserved global charges
and that the absence of global symmetry in quantum gravity
 \cite{Misner:1957mt, Banks:2010zn, Harlow:2018jwu, Harlow:2018tng} demands
their triviality. 

\smallskip

The cobordism conjecture predicts a domain wall for any pair of consistent quantum theories of gravity.
In particular, any quantum gravity must allow an end-of-the-world brane. 
In this paper, we discuss them in the context of the AdS/CFT correspondence. 
If there is a conformal interface between the corresponding pair of CFT's,
we can use it to construct such a domain wall. Similarly, a CFT with a conformally invariant boundary condition can be
dual to an AdS gravity with an end-of-the-world brane. Such end-of-the-world branes
and domain walls  have been discussed in the AdS/CFT correspondence 
starting with \cite{Randall:1999vf, Karch:2000ct, Takayanagi:2011zk} and \cite{DeWolfe:2001pq, Bachas:2001vj}, respectively.

\smallskip

However, there are topological constraints on possible conformal boundaries and interfaces \cite{Hellerman}. 
For example, in two dimensions, conformal invariance can be consistently imposed on an interface only when the gravitational anomaly
given by 
the difference $(c_L - c_R)$ of left and right Virasoro central charges matches across the interface.
 Let us write the left/right-moving Virasoro generators on the two side of the interface as $(L^{(1)}_n,\tilde{L}^{(1)}_n)$ and $(L^{(2)}_n,\tilde{L}^{(2)}_n)$, respectively.
A conformal interface must preserve their linear combination $(L^{(1)}_n-\tilde{L}^{(1)}_{-n} -L^{(2)}_{-n}+\tilde{L}^{(2)}_n)$ for all $n \in \mathbb{Z}$. Such an interface exists only when the central charge 
for this combination given by $(c^{(1)}_L  - c^{(1)}_R - c^{(2)}_L + c^{(2)}_R)$ vanishes.  
 We would like to see whether domain walls in AdS still exist when gravitational anomalies do not match, 
and if so in what sense. 

\smallskip

In order to address this question, we need to formulate properties of required domain walls more precisely. 
Consider a pair of quantum gravity theories in  $(d+1)$-dimensional AdS's, which we denote by AdS$_1$ and AdS$_2$.
According to the AdS/CFT correspondence, they are equivalent to conformal field theories 
in $d$ dimensions, which 
we call CFT$_1$ and CFT$_2$, respectively.
We say that there is a domain wall interpolating the two AdS gravity theories
 if there is a quantum field theory (which we denote by {\tt dw}QFT) 
on $\mathbb{R}_{{\rm time}} \times \mathbb{S}^{d-1}$
with the properties described in the following paragraph.

\smallskip

\bfig
\includegraphics[height=6cm]{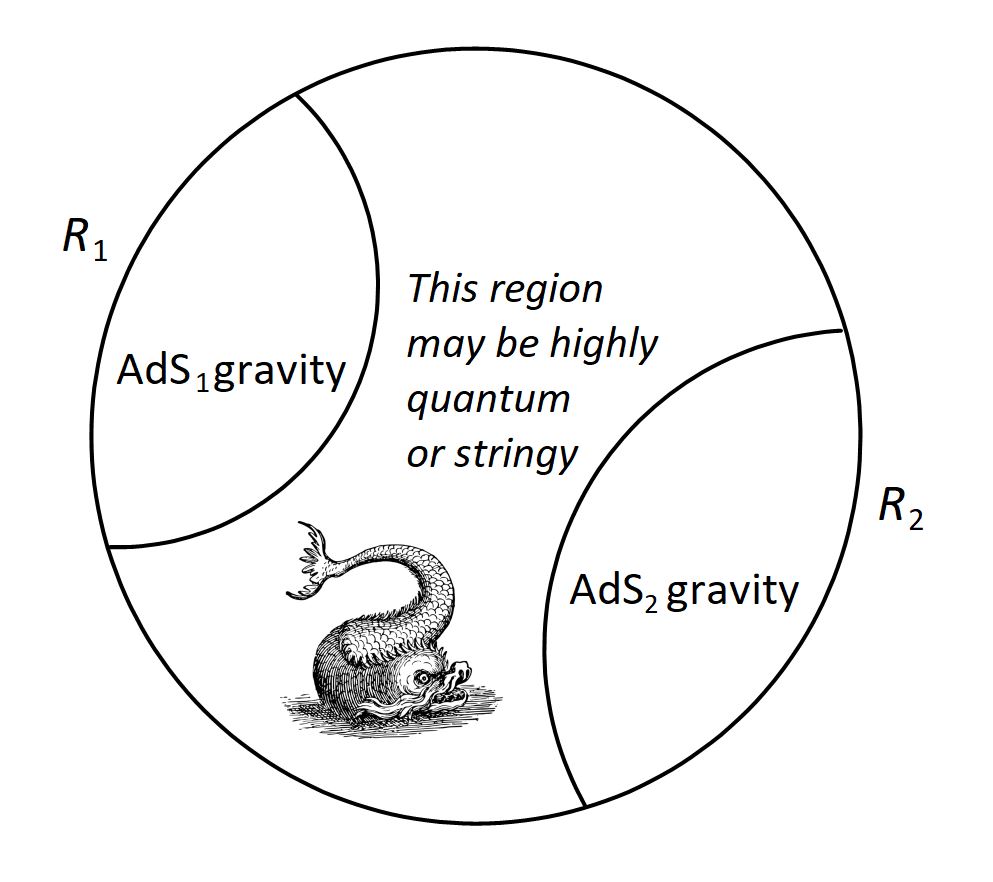}
\caption{{\tt dw}QFT contains subsectors described by weakly-coupled gravity
theories in subregions of AdS$_1$ and AdS$_2$. These subregions
extend all the way to the boundaries of AdS$_1$ and AdS$_2$, respectively.}\label{figone}
\efig

Though {\tt dw}QFT as a whole may not have a weakly-coupled gravity description in the bulk,
we require that it has subsectors which can be approximated arbitrarily precisely by the gravity theories in AdS$_1$ and AdS$_2$.
To formulate such a requirement mathematically,
consider contractible and mutually disjoint regions ${\cal R}_1$ and ${\cal R}_2$ on the Cauchy slice $\mathbb{S}^{d-1}$ of {\tt dw}QFT. 
We require that the algebra of local observables inside the
domain of dependence of ${\cal R}_a$ ($a=1,2$) can be described  approximately
by the AdS$_a$ gravity theory in the entanglement wedge $E_a$ for some region 
of the Cauchy slice of CFT$_a$. We also require that
the approximation can be made arbitrarily precise by making the region ${\cal R}_a$ small.
{\tt dw}QFT with these properties 
gives a quantum mechanical description of the domain wall predicted by the cobordism conjecture
since it interpolates 
the infinite-volume region $E_1$ 
described by the weakly coupled gravity 
theory in AdS$_1$ 
to the infinite-volume region $E_2$ described by the weakly coupled 
gravity theory in AdS$_2$, as shown in Figure \ref{figone}. 

\smallskip

If there is a conformal interface between CFT$_1$ and CFT$_2$, it naturally makes a dual to a domain wall
between AdS$_1$ and AdS$_2$ \cite{DeWolfe:2001pq, Bachas:2001vj}. 
To see that it satisfies the conditions stated in the above paragraph, 
note that CFT correlation functions sufficiently away from the interface can be
approximated by those without the interface and that the approximation can be made
arbitrarily precise by taking their insertion points close to each other and away from the interface. 
Thus, if we take ${\cal R}_a$ to be sufficiently small compare to its distance to the interface, 
its entanglement wedge is well-approximated by that of the original AdS$_a$ gravity. 

\smallskip

However, a conformal interface does not always exist. For example, 
M5 branes wrapping a 4-cycle $P_0$ of a Calabi-Yau manifold with the second
Chern class  $c_2 \cdot P_0$
is described at low energy by a two-dimensional CFT with the gravitational anomaly,
\begin{equation}
   c_L - c_R = - \frac{1}{2} c_2 \cdot q, 
\label{anomaly}
\end{equation}
where $q$ is the magnetic flux charge \cite{Kraus:2005zm}.
If CFT$_1$ and CFT$_2$ have different values of $c_2 \cdot q$, the conformal interface conditions cannot be
solved. Generally speaking, mismatch of gravitational anomalies signals difficulty in constructing an interface. 

\smallskip

Even if a conformal interface does not exist, it is still possible to connect  CFT$_1$ and CFT$_2$
at a junction of three conformal field theories, with an additional CFT$_3$ attached at the junction.
To make the junction possible, CFT$_3$ must carry gravitational anomalies that cancel those of CFT$_1$ and CFT$_2$. 
In the above
example, three sets of wrapped M5 branes
can join at a junction if their anomalies given by (\ref{anomaly}) add up to zero. 
For a reason we will discuss below, such a junction is always possible with an appropriate choice of CFT$_3$.
We can then stretch CFT$_3$ along the domain wall region between the AdS$_1$ and AdS$_2$ regions 
in Figure \ref{figone} and regard it as a part of degrees of the freedom of the region. 
Since  CFT$_3$ can couple to  CFT$_1$ and  CFT$_2$ consistently at the junction, it should also be 
possible place it between their bulk duals, $i.e.$,  AdS$_1$ and AdS$_2$.
Even when the  AdS$_1$ and AdS$_2$ gravities have different gravitational Chern-Simons terms generating
different amounts of anomaly inflows into the domain wall region, the mismatch of anomalies can 
be cancelled by having CFT$_3$ degrees of freedom in the region. 

\smallskip

If CFT$_3$ itself has large degrees of freedom and has an AdS dual, which we denote by AdS$_3$, 
it may be possible to construct a gravitational dual 
in terms of geometry with three asymptotically AdS regions as in \cite{Chiodaroli:2010mv}. In such a case, 
one can view that the AdS$_3$ region is emanating from the domain wall region in Figure \ref{figone}. 
The new AdS$_3$ branch can be regarded as an effective description of the degrees of freedom localized between 
the AdS$_1$ and AdS$_2$ gravities.

\smallskip

A similar construction can be considered for end-of-the-world branes predicted by the cobordism conjecture. 
If an AdS gravity carries gravitational Chern-Simons terms, the corresponding CFT has gravitational anomalies and
it cannot end on a conformal boundary. However, it is still possible to construct its conformal interface with 
another CFT with matching gravitational anomalies. This allows a construction of the predicted end-of-the-world brane
by bending the CFT toward the bulk and using it as degrees of freedom localized on the brane. The
anomaly inflow generated by the gravitational Chern-Simons terms can then be absorbed on the brane. 
One can even consider a totally transparent and topological interface, which connects CFT to itself. 
One can then bend the CFT on the other side of the interface and extend it toward the bulk AdS.
The resulting bulk geometry is the pure AdS with the Dirichlet boundary condition along the end-of-the-world brane.
This may not give a static configuration, but it is a consistent initial value condition on the Cauchy surface. 
This guarantees that there is an end-of-the-world brane for any AdS gravity. 
Similarly, there is a domain wall between any pair of AdS gravities since it can be regarded as
an end-of-the-world brane for the pair, by the standard folding trick for a conformal interface.
This shows that any potential topological obstruction against constructing a domain wall or an
end-of-the-world brane can be absorbed by an appropriate brane. 

\smallskip

A double Wick rotation of the configuration described in Figure \ref{figone} shows that  
the AdS$_1$ gravity can evolve into the AdS$_2$ gravity.
On the CFT side, the evolution is described by an interface operator which maps the Hilbert space of CFT$_1$ to 
that of CFT$_2$. If we require that sufficient information can be transmitted from
 AdS$_1$ to AdS$_2$, the rank of the interface operator must be large. 
However, it is not always the case, and the rank can
be as small as one. For example,
if both CFT$_1$ and CFT$_2$ allow conformal boundaries, one can consider an interface operator in the form of 
a tensor product of the boundary operator for CFT$_2$ and the hermitian conjugate of the boundary operator for
CFT$_1$. Such a totally reflective interface does not transmit information from  AdS$_1$ to AdS$_2$. 

\smallskip

In this paper, we discussed domain walls and end-of-the-world brane
 predicted by the cobordism conjecture in the context of the AdS/CFT correspondence.
We find that a domain wall between a pair of AdS gravities can exist even when gravitational anomalies of
the corresponding CFT's do not match and when there is no conformal interface between them,
by considering a junction of three CFT's so that the anomaly mismatch can be absorbed by 
the third CFT. The third CFT  can be regarded as a part of degrees of freedom of the domain wall. 
It may be necessary for the domain wall to carry large degrees of freedom, and their effective descriptions 
may involve new AdS branches emanating from the domain wall regions.

\bigskip
After completing this manuscript, we learned of work by Petar Simidzija and Mark Van Raamsdonk, who also
studied domain walls connecting different AdS gravity theories from the point of view of conformal interfaces \cite{MVA}. 

\medskip

\section*{acknowledgments}
{\small 
We thank Jacob McNamara, Cumrun Vafa, and Mark Van Raamsdonk for their comments on
this paper. We also thank Mark Van Raamsdonk for sharing
a draft of their paper \cite{MVA} prior to publication.
H.O. thanks participants of the
Bootstrapping String Theory Workshop of the Simons Collaboration
on the Nonpertubative Bootstrap for discussion.
The work of H.O. is supported in part by
U.S.\ Department of Energy grant DE-SC0011632, by
the World Premier International Research Center Initiative,
MEXT, Japan, by JSPS Grant-in-Aid for Scientific Research C-26400240,
and by JSPS Grant-in-Aid for Scientific Research on Innovative Areas
15H05895.
T.T. is supported by Inamori Research Institute for Science and 
World Premier International Research Center Initiative (WPI Initiative) 
from the Japan Ministry of Education, Culture, Sports, Science and Technology (MEXT). 
T.T. is also supported by the Simons Foundation through the ``It from Qubit'' collaboration, JSPS Grant-in-Aid for Scientific Research (A) No.16H02182 and 
by JSPS Grant-in-Aid for Challenging Research (Exploratory) 18K18766.
}


\bibliography{bibliography}
\end{document}